\documentclass[letterpaper,twocolumn,10pt]{article}
\usepackage{graphicx}
\graphicspath{{./figs}}
\usepackage{usenix,epsfig,bytefield}
\usepackage{ifthen}
\usepackage{xcolor}
\usepackage{subcaption}
\usepackage{booktabs}
\usepackage{color}
\usepackage{colortbl}
\usepackage{float}
\usepackage{bigdelim}

\newcommand{\showComments}{no}

\usepackage[hyphens]{url}
\usepackage[colorlinks,citecolor=blue,breaklinks=true]{hyperref}%
\usepackage{svg}
\usepackage{lipsum}
\usepackage{amsmath}
\usepackage{amssymb}
\usepackage{amsthm}
\usepackage{algorithm}
\usepackage{algorithmicx}
\usepackage{algpseudocode}
\usepackage{blindtext,tikz}
\usetikzlibrary{calc}
\usepackage{datetime2}

\begin{document}

\newcommand\hmm[1]{\ifnum\ifhmode\spacefactor\else2000\fi>1000
\uppercase{#1}\else#1\fi}

\newcommand{\System}{FAFO}
\newcommand{\SystemName}{Fast Ahead-of-Formation Optimization} 
\newcommand{\ParaLyze}{\textit{ParaLyze}}
\newcommand{\ParaBloom}{\textit{ParaBloom}}
\newcommand{\ParaFramer}{\textit{ParaFramer}}
\newcommand{\ParaScheduler}{\textit{ParaScheduler}}

\newcommand{\CompanyName}{LayerZero Labs Ltd}

\newcommand{\CostSavings}{91\%}
\newcommand{\LaptopTPS}{50K}
\newcommand{\OverSerial}{$84\times$}
\newcommand{\AbortRate}{1\%}
\newcommand{\TLPFromReordering}{$10\times$}

\newcommand{\note}[2]{
  \ifthenelse{\equal{\showComments}{yes}}{\textcolor{#1}{#2}}{}
}
\newcommand{\nt}[1]{\note{blue}{Note: #1}}
\newcommand{\ai}[1]{\note{pink}{AI-generated: To vet}}
\newcommand{\todo}[1]{\note{red}{TODO: #1}}
\newcommand{\daniel}[1]{\note{blue}{Daniel: #1}}
\newcommand{\isaac}[1]{\note{cyan}{Isaac: #1}}
\newcommand{\kui}[1]{\note{red}{Kui: #1}}
\newcommand{\thomas}[1]{\note{purple}{Thomas: #1}}
\newcommand{\raz}[1]{\note{pink}{Raz: #1}}
\newcommand{\step}[1]{\raisebox{.5pt}{\textcircled{\raisebox{-.9pt} {#1}}}}

\widowpenalty10000
\clubpenalty10000

\date{}

\title{\Large \bf \System{}: Over 1 million TPS on a single node running EVM while still Merkleizing every block}

\author{
    \normalfont{
        Ryan Zarick \hspace{0.7em}
        Isaac Zhang \hspace{0.7em}
        Daniel Wong \hspace{0.7em}
        Thomas Kim \hspace{0.7em}
        Bryan Pellegrino
    } \\
    \vspace{0.4em}
    \normalfont{
        Mignon Li \hspace{0.7em}
        Kelvin Wong
    }
    \\
  \CompanyName{}
}
\maketitle

\let\thefootnote\relax\footnotemark\footnotetext{Copyright \copyright{~2025
\CompanyName{}. All rights reserved.}}

\begin{abstract}
Current blockchain execution throughput is limited by data contention, reducing execution layer parallelism.
\SystemName{} (\System{}) is the first blockchain transaction scheduler to address this problem by reordering transactions \emph{before block formation} for maximum concurrency.
\System{} uses CPU-optimized cache-friendly Bloom filters to efficiently detect conflicts and schedule parallel transaction execution at high throughput and low overhead.

We integrate the Rust EVM client (REVM) into \System{} and achieve over 1.1 million native ETH transfers per second and over half a million ERC20 transfers per second on a single node (Table~\ref{table:headline}), with \CostSavings{} lower cost compared to state-of-the-art sharded execution.
Unlike many other existing high throughput blockchain execution clients, \System{} uses QMDB to Merkleize world state after \emph{every block}, enabling light clients and stateless validation for ZK-based vApps.

\System{} scales with minimal synchronization overhead, scaling linearly with additional CPU resources until it fully exploits the maximum parallelism of the underlying transaction flow.
\System{} proves that the high throughput necessary to support future decentralized applications can be achieved with a streamlined execution layer and innovations in blockchain transaction scheduler design.
\System{} is open-sourced at \url{https://github.com/LayerZero-Labs/fafo}.

\end{abstract}

\ifthenelse{\equal{\showComments}{yes}}{
\SetWatermarkText{CONFIDENTIAL}
\SetWatermarkScale{.5}
\tikz[overlay,remember picture]{
\node at ($(current page.west)+(0.4,0)$) [rotate=90] {\Huge\textcolor{gray}{\DTMnow}};}}{}

\section{Introduction}
Blockchain execution has historically traded off throughput and decentralization; higher throughput always meant higher capital requirements, reducing decentralization.
In theory this is asymptotically true, but most blockchains are at least partially bottlenecked by the grossly inefficient use of modern multi-core CPUs on the execution layer.
The goal of our work is to efficiently utilize multi-core CPUs to achieve high throughput at low cost without introducing the complexities of sharding.

At a high level, blockchain execution begins with a block \emph{producer} selecting transactions from the mempool and packing them into blocks; these blocks are executed against the blockchain state transition function (STF) by one or more nodes in a decentralized network.
In most blockchains, transactions are packed into blocks with little to no regard to parallel execution, and the execution client either runs these transactions in serial or tries to parallelize a potentially unparallelizable transaction flow.
As a result, the execution layer makes very inefficient use of modern multi-core CPUs, requiring expensive overprovisioning or sharding to achieve high transactions per second (TPS).

Blockchain transactions must achieve \emph{deterministic serializability}, meaning that the schedule of all transactions across all blocks must be equivalent to a deterministically chosen total ordering of those transactions.
In this paper, we focus on parallelizing execution of EVM~\cite{wood2014ethereum} transactions, but the techniques we leverage are applicable to most blockchain VMs.
We recognize that transaction serializability is achievable if the schedule of all \emph{storage operations} is \emph{conflict-serializable}~\cite{10.1145/360363.360369}, and that this property is sufficient to implement transactional semantics for most if not all blockchain VMs.
We take advantage of this in our end-to-end scheduling and execution runtime, \SystemName{} (\emph{\System{}}), which identifies and exploits transaction-level parallelism through heuristic transaction scheduling, dispatch, and execution.

\begin{table}[t]
    \centering
    \begin{tabular}{@{}l|r@{}}
        \textbf{Benchmark} & \textbf{Throughput (TPS)} \\
        \hline
        Native & 1,121,732 \\
        ERC20 & 565,956 \\
    \end{tabular}
    \caption{\System{} achieves up to 1.1\,M transactions per second (TPS) on a 96-core server using a synthetic benchmark (\S~\ref{sec:workload-tlp-analysis})}
    \label{table:headline}
\end{table}

\System{} not only exceeds the state-of-the-art for single-node throughput, but also \emph{Merkleizes every block}; this enables light clients and stateless validation which is required for Zero-Knowledge (ZK)-based vApps~\cite{zhang2025vapps}.
Many other blockchains~\cite{yakovenko2018solana,Sui} Merkleize at very low frequency or forego Merkleization altogether to sidestep the storage bottleneck; \System{} instead leverages the QMDB verifiable database~\cite{zhang2025qmdb} to Merkleize at high throughput with no compromises.

Our contributions in this paper are threefold: first, a novel CPU-efficient, cache-friendly data structure (\ParaBloom{}) to detect data conflicts.
Second, an efficient algorithm (\ParaFramer{}) that uses \ParaBloom{} to generate a low-contention, highly parallelizable transaction stream.
Third, we design an efficient scheduler (\ParaScheduler{}) that schedules conflict-serializable parallel execution of the transaction stream.
These three components are combined into an efficient end-to-end scheduler and execution pipeline (\System{}), which we demonstrate can process over 1.1 million TPS and Merkleize every block \emph{on a single node} (Table~\ref{table:headline}) running EVM transactions on REVM backed by QMDB.

\section{Design}
In this section, we present the design of \System{} in detail and argue its correctness and optimality.
We define conflict-serializability and transaction-level parallelism in the context of blockchain execution (\S\ref{sec:tlp-concept}).
In \S\ref{sec:design:architecture}, we present a high-level overview of the four-stage pipelined design of \System{}.
We follow this with a detailed description of each component: \ParaLyze{} (\S\ref{sec:paralyze}), \ParaFramer{} \& \ParaBloom{} (\S\ref{sec:parabloom}), and \ParaScheduler{} (\S\ref{sec:runtime}).

\subsection{Transaction-level parallelism}
\label{sec:tlp-concept}
Two transactions \emph{conflict} if they both access the same data and at least one access is a write, and a \emph{conflict-serializable} transaction schedule $S'$ is such that there exists a serial schedule $S$ where the result of all conflicting memory accesses is equal between $S$ and $S'$ ~\cite{10.1145/360363.360369}.

Parallelism exists on a \emph{per-transaction} granularity: transactions can run concurrently as long as their storage accesses are conflict-serializable.
To capture this idea, we define Transaction-Level Parallelism (TLP) as the average number of transactions that can be run concurrently given some scheduling of a set of transactions.
There exists a theoretical upper bound on TLP for any set of transactions, and we refer to the \emph{optimal schedule} as any transaction schedule that achieves this upper bound.

The optimal schedule can be constructed via an optimal coloring of a precedence graph across all transactions, and real-world performance is maximized if the size of each color class is roughly equal to the number of execution units (cores) of the execution client.
Unfortunately, it is impractical to compute the globally optimal coloring for two reasons.
First, the computational cost of constructing and coloring the graph is high.
Second, the streaming nature of the workload demands some bound on service latency, limiting the number of unexecuted transactions that can be accumulated in the mempool.
\System{} strikes a balance between theory and practice, extracting the majority of parallelism from the workload without degrading end-to-end latency or overusing compute resources.

\subsection{\System{} Architecture}
\label{sec:design:architecture}

\begin{figure}
\centering
\includegraphics[width=\linewidth]{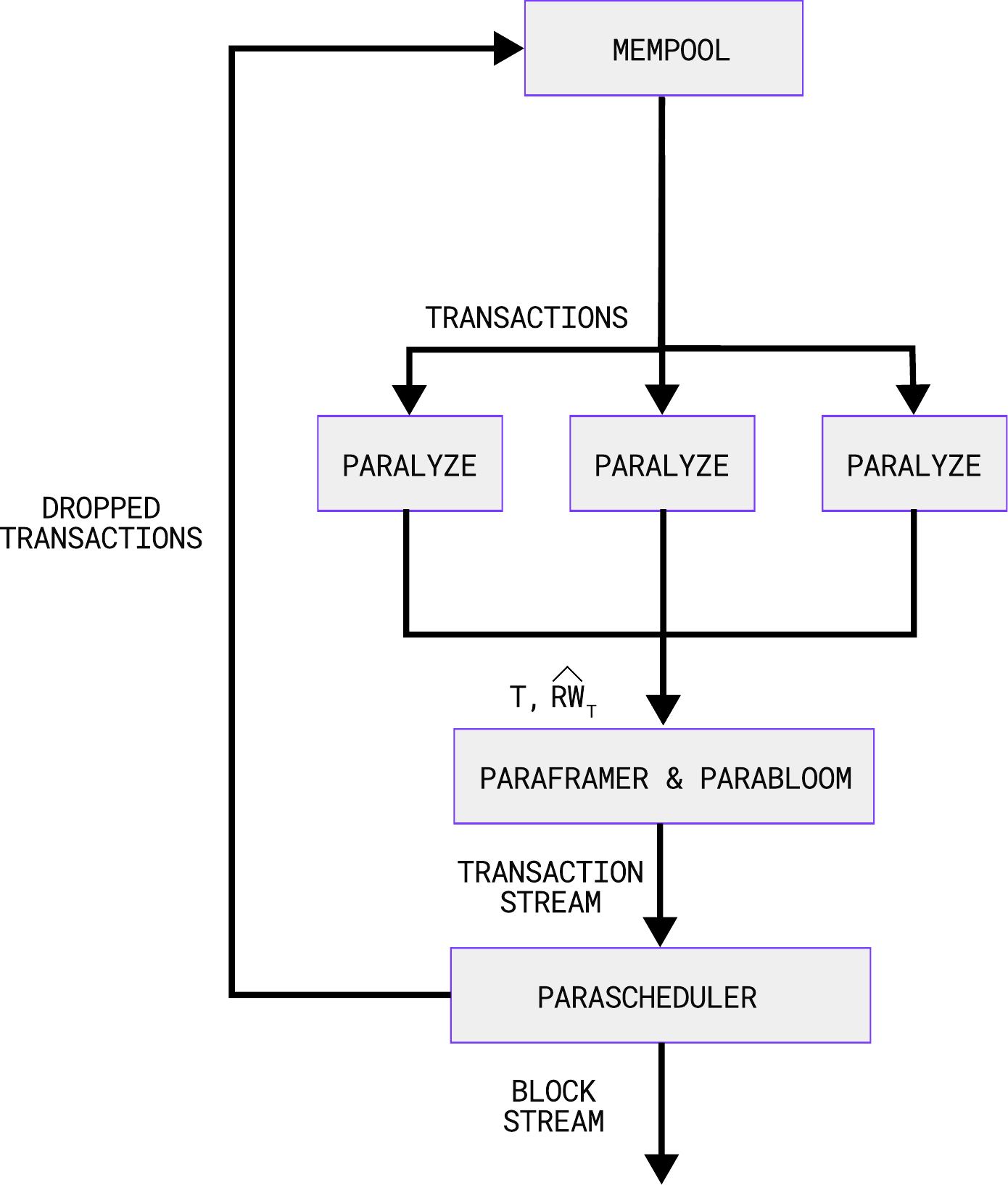}
\caption{\textbf{System Architecture.} Concurrent instances of \ParaLyze{} preprocesses transactions, \ParaFramer{} packs frames using \ParaBloom{}, and \ParaScheduler{} produces the final transaction execution ordering and defines block boundaries.}
\label{fig:arch}
\end{figure}

We assume that block producer election occurs on a reasonably long timescale (e.g., EOS~\cite{eoswhitepaper2018}, Solana\cite{yakovenko2018solana}) to increase the maximum TLP of the pool of transactions available to a single block producer.
\System{} is compatible with a wide range of blockchain applications such as vApps~\cite{zhang2025vapps}, rollups, and L1 chains.
In our model, the block producer produces a single totally-ordered stream of transactions $\{T_0, T_1, \dots, T_i\}$ and inserts block headers into the stream to mark block boundaries.
Without loss of generality, we assume that the underlying verifiable database (e.g., QMDB~\cite{zhang2025qmdb}) is not a bottleneck of the system.

\System{} processes transactions in a four-stage pipeline:

\begin{enumerate}

\item \textbf{\ParaLyze{}} preprocesses each transaction in the mempool to approximate the addresses of every storage slot it reads and writes.
This approximation is \emph{not} guaranteed to accurately reflect the state accessed during execution, as the ordering of transactions is not final at this stage.

\item \textbf{\ParaFramer{}} identifies non-conflicting transactions by computing the intersection of their approximate read/write sets using \ParaBloom{} (\S\ref{sec:parabloom}).
It then packs these non-conflicting transactions into \emph{frames}, and outputs a stream of transactions grouped by frame.

\item \textbf{\ParaScheduler{}} dispatches and executes the stream of transactions received from \ParaFramer{}, identifying and exploiting any transaction-level parallelism in the process.
Transactions whose actual read/write set diverge from their approximate read/write sets are dropped and returned to the mempool.

\item \textbf{Block formation}: \System{} periodically synchronizes the execution threads, flushes storage, and inserts block headers into the stream of non-dropped transactions output by \ParaScheduler{}.
These blocks can then be settled to the underlying consensus layer.

\end{enumerate}

\subsection{\ParaLyze{}: Transaction Static Analysis}
\label{sec:paralyze}

\ParaLyze{} enforces that all transactions in the mempool are well-formed and annotated with the set storage reads and writes that they will perform.
We term this the \emph{approximate read/write set} and denote the approximate read/write set of a transaction $T_i$ as $\widehat{RW}_{T_i} = \{ \widehat{R}_{T_i}, \widehat{W}_{T_i} \}$.
$\widehat{RW}_{T_i}$ is determined either by metadata (e.g., full node-provided access list~\cite{eip-2930}) or simulation; without loss of generality, we assume the latter.

\ParaLyze{} does not persist any writes during simulation to avoid bottlenecking this stage on storage contention; as a result, many transactions are simulated against the same initial state which may diverge (in a conflicting manner) from the state during actual execution.
We use distributed parallel simulation to maximize throughput, and accept that approximate read/write sets may sometimes be inaccurately captured for workloads such as data-dependent flows or those which rely heavily on modifying global state.

\subsection{\ParaFramer{} \& \ParaBloom{}: Conflict detection and frame packing}
\label{sec:paraframer}

\begin{figure}[t]
    \centering
    \includegraphics[width=\linewidth]{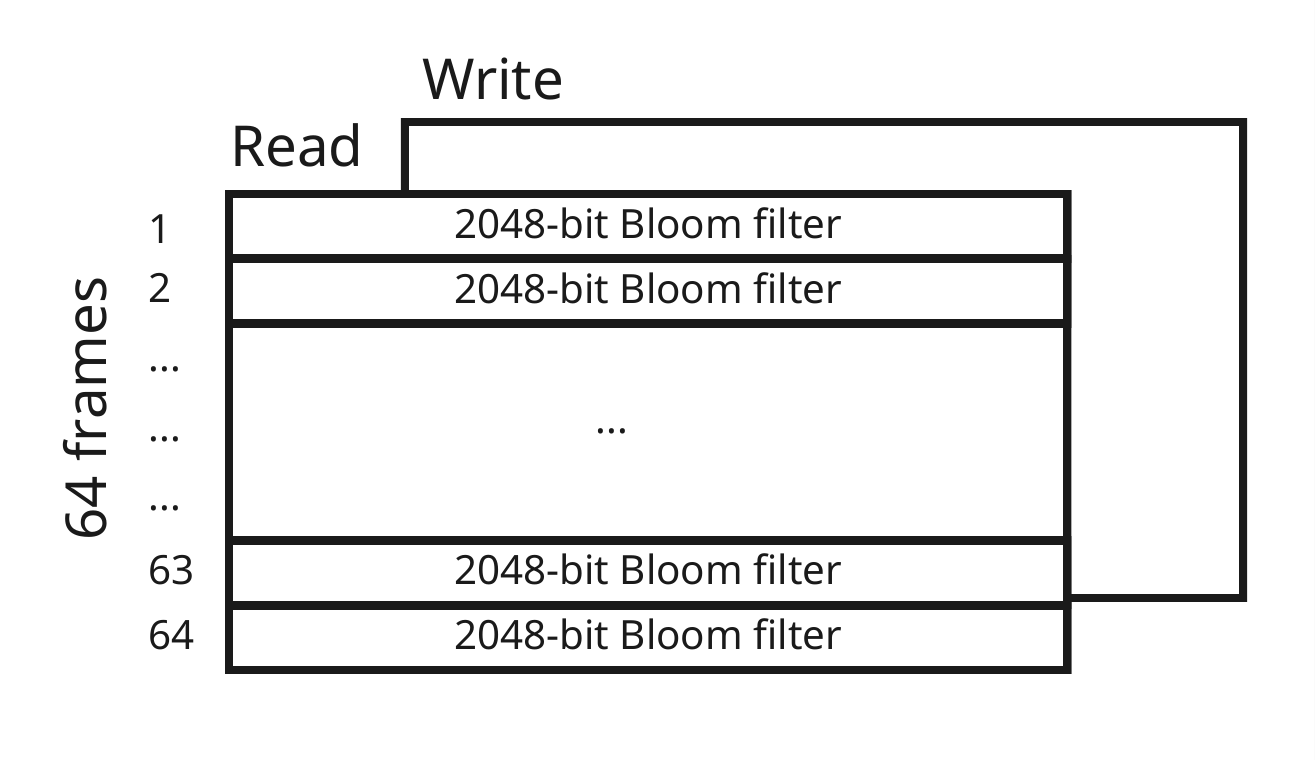}
    \caption{\textbf{\ParaBloom{} Layout.} Each active frame stores two 2048-bit Bloom filters (read and write).}
    \label{fig:bloom-filter-appendix}
\end{figure}

\ParaBloom{} identifies and groups non-conflicting transactions into \emph{frames}, and \ParaFramer{} forms the transaction stream from these frames.
This transaction stream is \emph{parallelism-aware}, with average TLP greater than or equal to the average number of transactions per frame.

We implement \ParaBloom{} using a collection of Bloom filters to efficiently detect data conflicts.
\ParaBloom{} compactly represents the union of the read and write sets of all transactions in each frame using two Bloom filters, one for read and one for write.
This allows efficient conflict checks via CPU-efficient bitwise operations.

\ParaFramer{} does not generate a \emph{theoretically} optimal transaction stream due to the two impracticalities mentioned at the end of \S\ref{sec:tlp-concept}.
First, the Bloom filters in \ParaBloom{} have some probability of falsely identifying data conflicts; this reduces TLP by about 8\% in our testing but also significantly improves the end-to-end throughput and efficiency of \System{}.
Second, we do not perform an exhaustive search of all combinations of transactions in the mempool, but instead greedily pack them in a single streaming pass over the mempool.

The computational complexity of conflict-serializable scheduling is generally linear in the size of the precedence graph~\cite{cellary2014concurrency}, so a high degree of contention increases the number of edges and, by extension, the computational cost of scheduling.
Our frame-based approach uses lightweight cache-friendly data structures and approximation to amortize the cost of checking conflicts and reduce the number of checks respectively.

\paragraph{\ParaBloom{}:}
\label{sec:parabloom}

For transaction $T$ packed in frame $F_i$, we store $\widehat{RW}_T$ in \ParaBloom{} for frame $F$ ($PB_{F}$).
Each frame in \ParaBloom{} is represented by two Bloom filters: one for the frame's \emph{aggregate read set} ($\widehat{AR}_{F}$), and another for its \emph{aggregate write set} ($\widehat{AW}_{F})$ (Figure~\ref{fig:bloom-filter-appendix}).
Transaction $T$ is \emph{admissible} to frame $F_i$ iff no storage operation conflicts between $F_i$ and $T$, or more formally:

\begin{equation}
    \label{eq:parabloom}
    \begin{aligned}
        Adm&issible(T, \widehat{RW}_{T}, \widehat{PB}_{F}): \\
        & (\widehat{R}_{T} \cap \widehat{AW}_{F}) = (\widehat{W}_T \cap \widehat{AR}_{F}) = (\widehat{W}_T \cap \widehat{AW}_{F}) = \varnothing
    \end{aligned}
\end{equation}

\paragraph{\ParaBloom{} Frame Packing:}

\begin{algorithm}[H]
\caption{Greedy Frame-Packing Algorithm}
\label{alg:greedyframepacking}
    \begin{algorithmic}[1]
        \State \textbf{Input:} Transaction list $T$, read/write sets $RW_{T_i}$
        \State \textbf{Output:} Frames (sets of non-conflicting txns)
        
        \State Initialize $F$ empty frames
        \For{$T_j$ in $T$}
            \State \texttt{placed} $\gets \texttt{false}$
        \For{each frame $F_i$ in \texttt{Frames}}
            \If{Admissible($t, RW_{T_i}, PB_{F_i}$) \text{(Eq.\,(\ref{eq:parabloom}))}}
                \State $txns_i \gets txns_i \cup \{t\}$ 
                \State $\widehat{AR}_{F_i} \gets \widehat{AR}_{F_i} \cup \widehat{R}_{T_j}$
                \State $\widehat{AW}_{F_i} \gets \widehat{AW}_{F_i} \cup \widehat{W}_{t_j}$
                \State \texttt{placed} $\gets \texttt{true}$; \textbf{break}
            \EndIf
        \EndFor
        \If{\texttt{placed} = \texttt{false}}
            \State $f_x \gets $ largest frame
            \State $Finalize(f_x)$
            \State $f_x \gets \{T_j\}$
            \State $\widehat{AR}_{T_j} \gets \widehat{R}_{T_j}$
            \State $\widehat{AW}_{T_j} \gets \widehat{W}_{T_j}$
        \EndIf
        \EndFor
    \end{algorithmic}
\end{algorithm}

\ParaBloom{} packs frames greedily, placing each transaction into the lowest index non-conflicting frame in \ParaBloom{}.
If the transaction conflicts with all frames in the active set, \ParaBloom{} finalizes (commit) the largest existing frame, replaces it with a new empty frame, then inserts the new transaction into the empty frame (illustrated in Algorithm~\ref{alg:greedyframepacking}).
This design encourages large average frame sizes, directly correlated with a better lower bound of TLP.
These ejected frames are returned to \ParaFramer{} which then sends them to \ParaScheduler{}.
Optionally, \ParaFramer{} can define a secondary policy (e.g., number of transactions, time) to eject frames from \ParaBloom{}.

\ParaBloom{} has two parameters: the number and size of each Bloom filter.
Our testbed uses a 64-bit CPU with 64\,kB of L1 cache, so we configure \ParaBloom{} to use 64 pairs of 2048-bit Bloom filters; this allows \ParaBloom{} to index frames using a bitmap stored in a single 64-bit word  and fit neatly into 32\,KiB (half of L1 cache).
We recommend parameterizing the number of frames based on the word size of the CPU, then sizing each Bloom filter with the maximum number of bits that will keep the entire structure within half of L1 cache.

\subsection{\ParaScheduler{}: Transaction Scheduling and Execution} \label{sec:runtime}

\ParaScheduler{} begins by reconstructing the same frames generated by \ParaFramer{} from the transaction stream.
As frames are ejected from the active set, we insert each of its transactions $T_i$ into a collection of directed acyclic graphs (DAGs).
For each storage slot $s$ in $RW_{T_i}$, there exists a corresponding DAG; the transaction is inserted into the DAG and a directed edge is added from each transaction $T_j$ that conflicts with $T_i$ such that ($T_j \xrightarrow{} T_i \land \lnot \exists T_k | T_j \xrightarrow{} T_k \xrightarrow{} T_i, j < k$ where $\xrightarrow{}$ is the happens-before relation).
Each of the DAGs is constructed by a separate thread, and executed transactions are pruned from the DAG to reduce the iteration space.
Furthermore, $T_i$ does not need to be checked against any other transactions in the same frame, as by construction there are no data conflicts.

\ParaScheduler{} dispatches a transaction immediately after all ancestors have completed execution to ensure that transactions are safely dispatched and executed as soon as possible.
After execution, the actual read/write set of the transaction $T_i$ ($RW_{T_i}$) is recorded and compared to $\widehat{RW}_{T_i}$.
If $RW_{T_i} \neq \widehat{RW}_{T_i}$, $T_i$ is dropped from the stream to preserve conflict-serializability, then returned to the mempool to be rescheduled; we expect this to be rare in practice.
Intuitively, the TLP of the transaction schedule executed by \ParaScheduler{} will be equal to or greater than $\frac{transactions}{frames}$, since \ParaScheduler{} can at least execute all transactions in each frame in parallel.

We argue that this collection of DAGs is equivalent to a precedence graph, and that the schedule produced by this DAG achieves equal or greater TLP than executing the transactions frame-by-frame.
Without loss of generality, we assume an unbounded number of CPU cores on the server running \ParaScheduler{}.
Suppose only 1 storage slot is ever accessed: by construction, the DAG produced by \ParaScheduler{} is equivalent to the precedence graph of the operations.
Additionally, the total ordering of the stream ensures this DAG is a forest of directed trees.
Under the same assumption, there must exist no edge between any two transactions in the same frame; \ParaScheduler{} only delays execution of any given transaction $T_i$ if its direct ancestor $T_j$ has not yet been executed, and $T_i$ must have no direct ancestor within the same frame.

Relaxing the assumption of a single storage slot, the ``schedule'' $S$ generated by the DAG for each storage slot $a$ (denoted $S_{a}$) is conflict-serializable for that particular slot.
Each transaction is executed only after the schedule $S_{a}$ permits, so it is impossible for the executed schedule to have any conflicts for slot $a$.
For a collection of DAGs, \ParaScheduler{} waits until all schedules $S_{a_x}$ permit the execution of transaction $T_i$ before dispatching $T_i$, guaranteeing that the schedule of operations in $T_i$ is conflict-serializable across all storage slots $S_{a_x}$.

\paragraph{Summary}
\System{} reorders transactions to maximize TLP \emph{ahead of} block formation--this design choice enables \System{} to use an optimistic pipeline for high throughput.
\ParaLyze{} optimistically precomputes read/write sets, which are used by \ParaFramer{} to generate transaction streams with high TLP.
\ParaFramer{} uses \ParaBloom{} to detect conflicts in a cache-friendly CPU-efficient manner, and amortizes conflict checks across multiple transactions using \emph{frames}.
\ParaScheduler{} uses frames to efficiently compute and execute the optimal conflict-serializable schedule of the transaction stream, identifying opportunities for concurrent execution of transactions within and across frames.

\section{Evaluation}
\label{sec:evaluation}

We demonstrate \System{}'s high throughput and scalability even on workloads with large state sizes and high contention.

\subsection{Setup}
\label{sec:setup}

\paragraph{Hardware}
All benchmarks run on a single AWS \texttt{i8g.metal-24xl} instance with 96 vCPUs (ARM-based Graviton3 cores), 768\,GiB of DRAM, and 6 local NVMe SSDs in a RAID0 configuration (aggregated 22.5\,TB, 2.16\,M IOPS, 26\,GB/s read and 20\,GB/s write).

\subsection{Workloads}
\label{sec:workload-tlp-analysis}

We evaluate \System{} using two synthetic workloads: \emph{Native} transfer and \emph{ERC20} transfer.
\emph{Native} transfers ETH from one account to another, changing two state slots (the sender balance and receiver balance). 
\emph{ERC20} transfers an ERC20 token from one holder to another, changing three state slots (the sender ETH balance, the sender token balance, and the receiver token balance).
We only run \ParaLyze{} on a subset of the ERC20 transfers, to simulate the behavior of the system if EIP-2930~\cite{eip-2930} is implemented.

We generate batches of transfers parameterized by \emph{skew} and \emph{contention ratio}, described below.
\begin{itemize}
  \item \textbf{$\gamma$ (Skew):} The number of \emph{hot} addresses that receive transfers more frequently. Lower $\gamma$ exacerbates storage hotspots and reduces maximum TLP.
  \item \textbf{$\alpha$ (Contention Ratio):} The probability of transferring to a hot address.
\end{itemize}
We represent the specific benchmark parameterization of workload (e.g., ERC20), skew Y, and contention ratio A as $ERC20_{\alpha=A, \gamma=Y}$ (if $\alpha = 0$, we omit $\gamma$ for brevity).

The \texttt{from} address is uniformly randomly chosen from the set of \textit{hot} addresses with probability $\alpha$, and chosen from the remainder of the addresses with probability $1 - \alpha$.
The \texttt{to} address is always uniformly randomly chosen from the non-hot addresses.
Hot keys are chosen (uniformly) from a narrow, contiguous range of size $\gamma$ from the beginning of the lexicographically sorted list of addresses.

\begin{figure}[t]
  \centering
    \includegraphics[width=\linewidth]{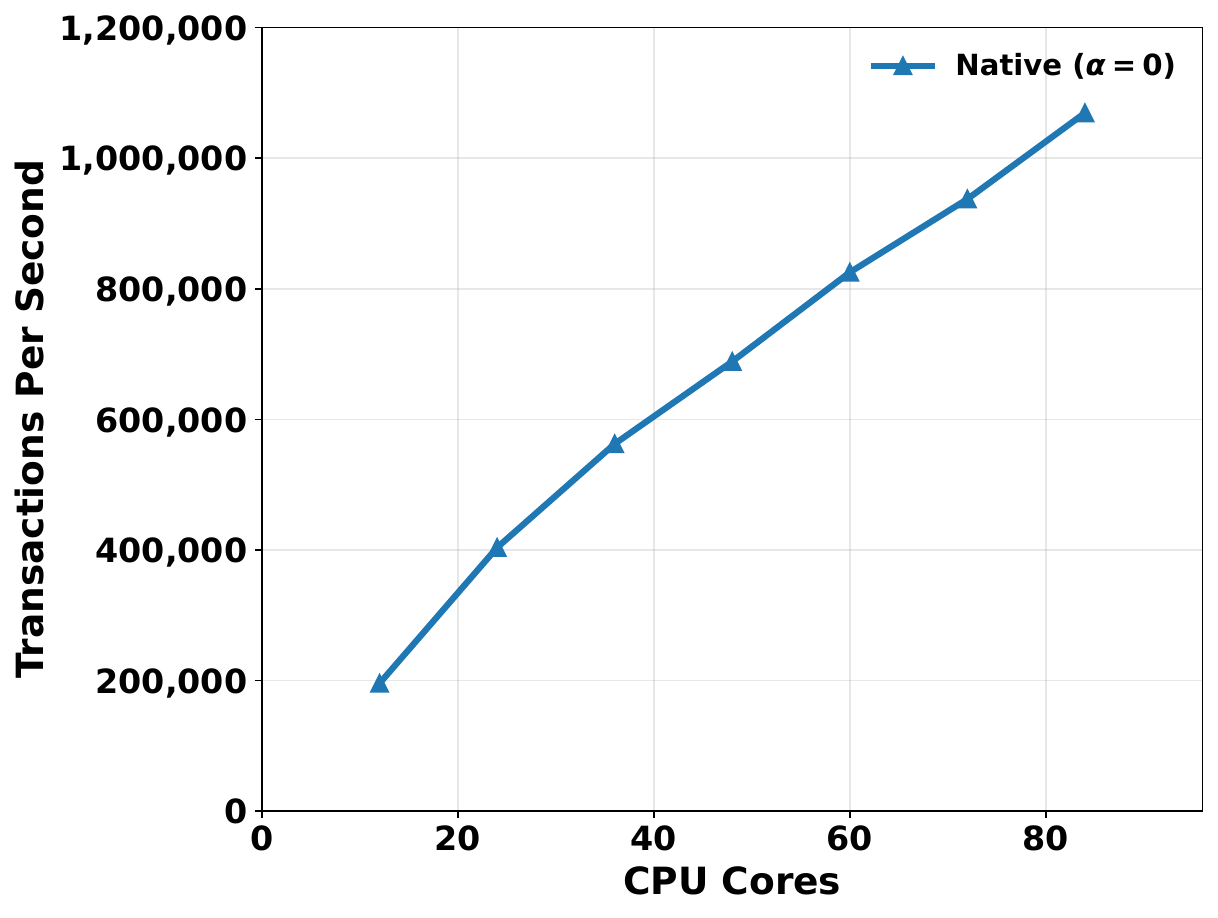}
    \caption{\textbf{\System{} scales linearly.} \System{} extracts about 130 TLP from $native_{\alpha=0}$, theoretically allowing it to scale up to 130 CPU cores. \System{} efficiently uses each additional CPU core, up to the maximum available (96).}
  \label{fig:throughput-scaling}
\end{figure}

\begin{figure*}[t]
    \centering
    \includegraphics[width=\linewidth]{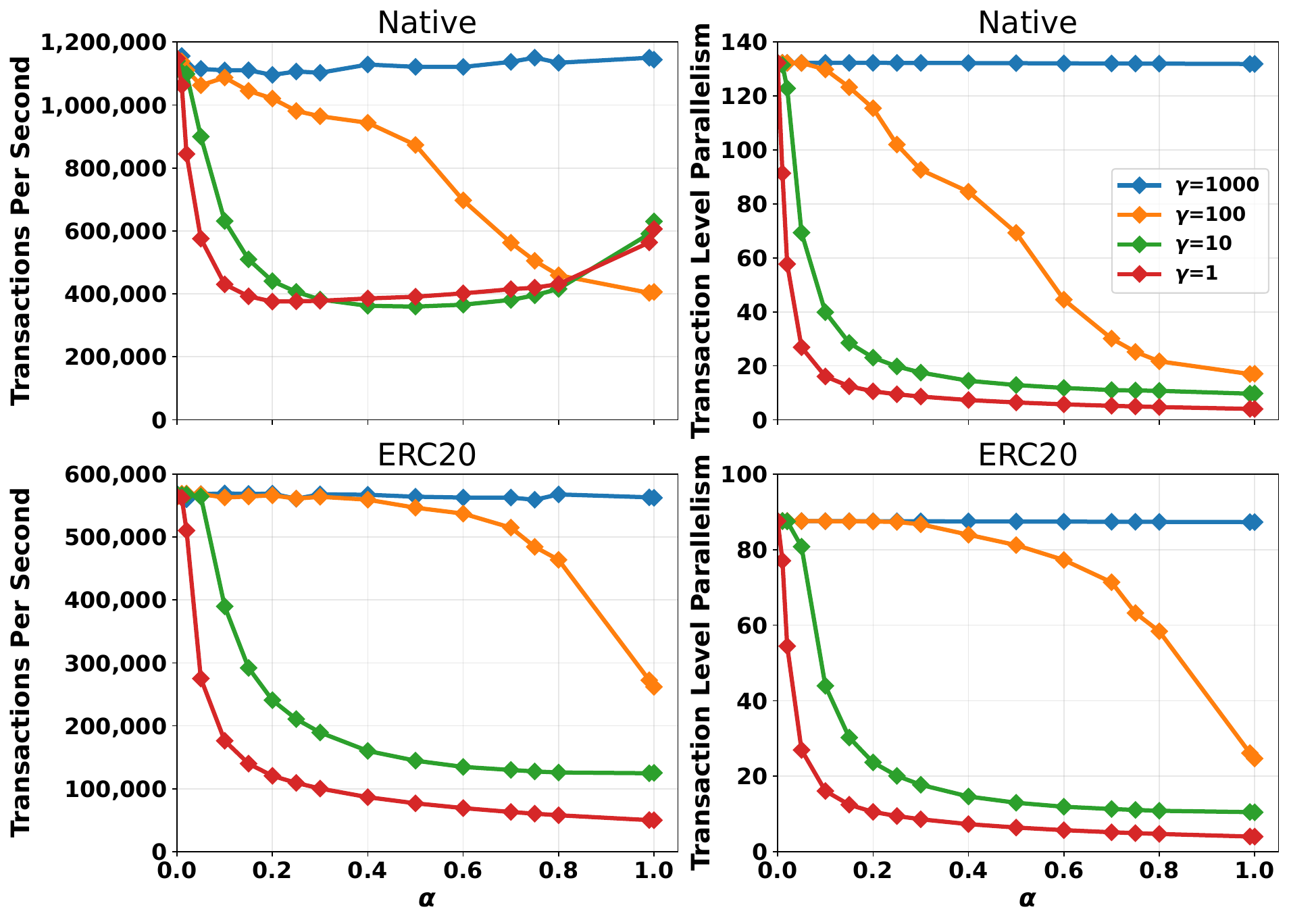}
      \caption{We run \System{} under varying combinations of $\alpha$ and $\gamma$ (\S~\ref{sec:workload-tlp-analysis}) on a total dataset size of $2^{30}$ keys.
      Higher contention reduces TLP and TPS, but FAFO achieves over 1.1 million TPS even with 99\% of requests hitting only 1000 keys (approx one per million).}
      \label{fig:contention}
\end{figure*}

In all benchmarks, we pre-populate the state with $2^{30}$ (approx. 1~billion) accounts, then issue 500\,K batched transfers on each of 512 threads (256\,M transactions total).
We then issue 512 additional \textit{transfer blocks}, each with approximately 500,000 batched transfers.
For both \emph{native} and \emph{ERC20}, each transfer is represented as a 24-byte record (6~bytes sender address, 6~bytes receiver address, 12 bytes of offsets), with the address of the global ERC20 contract being implicit.

\subsection{Results}
\label{sec:eval:throughput-scaling}

\paragraph{\System{} scales with additional CPU cores}
\label{sec:eval:contention}
Fig~\ref{fig:throughput-scaling} shows that \System{} throughput scales \textbf{linearly} with each additional CPU core available; this trend is expected to plateau when the number of cores exceeds the TLP of the underlying workload.

\paragraph{\System{} is \CostSavings{} cheaper than sharding.}
\System{} achieves over 1 million TPS at less than 10\% of the cost of the state-of-the-art sharding-based approach (approximately 6,013 vs 65,361 USD per month based on monthly on-demand pricing for AWS~\cite{aws-ondemand-pricing} and GCP~\cite{gcp-ondemand-pricing} at the time of writing).

\paragraph{\System{} handles contention well} We run the \emph{transfer} workload with varying values of $\alpha$ and $\gamma$, measuring the TLP and TPS of \System{} (Figure~\ref{fig:contention}.
Even under unrealistically adversarial conditions (99\% of requests accessing 0.0001\% of the storage slots), \System{} maintains over 130 TLP and 1.1 million TPS.
We believe that real-world workloads will be much less skewed, with real blockchain hotspots only being 0.1\% of storage slots accounting for 62\% of accesses (Section~\ref{sec:related-work}).
This 0.1\% of storage slots corresponds to $\gamma > 1,000,000$, which is omitted from our results as it is indistinguishable from $\gamma = 1000$.
The high throughput of \System{} coupled with the massive state sizes enabled by QMDB makes it unrealistic for any real workload to have fewer than 10,000 or even 100,000 hot storage slots.

The unusually high TPS of \System{} for small values of $\gamma$ and large values of $\alpha$ is likely due to extremely high cache locality of the execution client workload when the context for all hot accounts fit into L1 cache.

For the \emph{ERC20} workload, the throughput of \System{} drops end-to-end by about 50\%; this is because each ERC20 transfer writes 50\% more state (3 entries vs 2 entries) and requires more computation per transaction (contract vs native).

\paragraph{\System{} has low CPU overhead}
\label{sec:eval:microbenchmarks}
We perform an ablation study to compare the CPU overhead of \System{}.
Our benchmark with only \ParaFramer{} and \ParaScheduler{} (skipping transaction execution) schedules over 2 million TPS.

\section{Related Work}
\label{sec:related-work}

\paragraph{Increasing execution layer parallelism}
A challenge facing modern blockchains is the \emph{hotspot problem}, where 0.1\% of storage slots account for 62\% of
accesses~\cite{lin2025parallelevm}, resulting in high contention that limits parallelism. This has limited the efficacy of
existing approaches that rely on optimistic concurrency control (OCC)
to modest speedups of just $2\mbox{--}5\times$ on EVM workloads ~\cite{shahid2024parallel,lin2025parallelevm}.
Sharding approaches~\cite{aptos2025shardines} have seen some success scaling horizontally, but introduce new synchronization and storage bottlenecks (exacerbated by the hot spot problem).

Shardines~\cite{aptos2025shardines} scales Block-STM but runs into storage bottlenecks at 30 nodes and observes sublinear scaling and diminishing marginal returns per additional shard with a 33\% drop in efficiency for workloads with contention when the number of shards was doubled.
In our work, we demonstrate that it is possible to achieve more than enough throughput to serve current blockchain applications without introducing the complexity or cost of sharding.
However, sharding is an orthogonal approach to \System{}, and these approaches can be composed if necessary.

\paragraph{Parallel Execution in EVM Blockchains.}
Prior attempts at accelerating Ethereum's sequential execution rely on optimistic concurrency control (OCC) and speculative execution.
Block-STM~\cite{gelashvili2023block} applies multi-version OCC to speculatively run transactions in parallel, which boosts throughput but has been shown to suffer from high abort overhead under high contention~\cite{lin2025parallelevm}.
ParallelEVM~\cite{lin2025parallelevm} opts for more fine-grained operation-level concurrency and uses a static single-assignment (SSA) log to trace operation dependencies, identify conflicts and re-execute, with a reported speedup of $4.28\times$.
Forerunner~\cite{chen2021forerunner} speculatively executes transactions in the window of time between dissemination and executes multiple possible futures in parallel execution, using memoization to speed it up.

\paragraph{Transaction reordering}
Replica-side schedulers such as OptME~\cite{ryu2024toward} and DMVCC~\cite{qi2023smart} reorder \emph{statements} after the block is disseminated.
Because every replica must reproduce the identical schedule, they either (i) speculatively execute each transaction to discover its read/write set (high overhead) or (ii) require users to supply access lists (poor UX).
They also cannot embed proposer-side policies (anti-spam, local fee markets), and any algorithmic change requires a hard fork.
In contrast, \System{} block producers reorder \emph{transactions} before block formation, reducing per-validator CPU overhead and enabling arbitrary scheduler upgrades.
Hyperledger-style systems (e.g., Fabric++~\cite{sharma2019blurring}, FabricSharp~\cite{ruan2020transactional}, and HTFabric~\cite{song2024htfabric}) reorder to minimize aborts inside an Execute-Order-Validate model, but this EOV model is orthogonal to the Order–Execute architecture of account-based blockchains.

\paragraph{Balancing static analysis and speculative execution}
Solana~\cite{yakovenko2018solana} avoids aborts entirely through lock-based scheduling on static read/write sets, but offloads the burden of accurate resource specification to contract developers.
At the other extreme, Dickerson \emph{et al.}~\cite{dickerson2017adding} and OptSmart~\cite{anjana2024optsmart} propose speculative scheduling on the block producer; this lowers validator conflicts in exchange for increased CPU overhead on the leader.
\System{}'s hybrid approach strikes a better balance between these two extremes.

\section{Conclusion} We introduce \System{}, a transaction scheduler that schedules and executes over 1.1 million EVM transactions per second on a single node .
\System{} delivers near-linear throughput scaling with CPU core count and achieves the same throughput as sharded systems with \CostSavings{} lower cost.

\System{} departs from speculation-reliant models by reordering transactions \textit{before} block formation to better exploit opportunities for parallelism.
This maximizes concurrency and minimizes synchronization overhead by combining efficient approximate conflict detection (\ParaBloom{}) with lightweight static scheduling (\ParaScheduler{}).

By reducing the cost of high-throughput execution and lowering capital costs, \System{} enables high throughput without sacrificing decentralization.

{\footnotesize
\bibliographystyle{acm}
\bibliography{bib}}

\end{document}